\documentclass[a4paper,usenatbib]{mnras}
\pdfoutput=1

\usepackage[T1]{fontenc}
\usepackage{ae,aecompl}

\usepackage{graphicx}	
\usepackage{amsmath}	
\usepackage{amssymb}	
\usepackage[flushleft]{threeparttable}

\title[Charactering the M83 companion dw1335-29]{Characterizing dw1335-29, a recently discovered dwarf satellite of M83 }

\author[A. Carrillo et al.]{Andreia Carrillo$^{1}$,
Eric F.\ Bell$^{1}$,
Jeremy Bailin$^{2}$,
Antonela Monachesi$^{3}$,
\newauthor Roelof S.\ de Jong$^{4}$, 
Benjamin Harmsen$^{1}$,
and Colin T.\ Slater$^{5}$
\\
$^{1}$Department of Astronomy, University of Michigan, 311 West Hall, 1085 S.\ University, Ave., Ann Arbor MI 48109, USA\\
$^{2}$Department of Physics and Astronomy, University of Alabama, Box 870324, Tuscaloosa, AL 35487-0324, USA \\
$^{3}$Max Planck Institute for Astrophysics, Karl-Schwarzschild-Str. 1, Garching D-85748, Germany\\
$^{4}$Leibniz-Institut f\"ur Astrophysik Potsdam (AIP), An der Sternwarte 16, 14482 Potsdam, Germany \\
$^{5}$Department of Astronomy, University of Washington, Box
351580, Seattle, WA 98195, USA }
\date{DRAFT : \today}

\pubyear{2016}

\begin{document}
\label{firstpage}
\pagerange{\pageref{firstpage}--\pageref{lastpage}}
\maketitle

\begin{abstract}
The number, distribution, and properties of dwarf satellites are crucial probes of the physics of galaxy formation at low masses and the response of satellite galaxies to the tidal and gas dynamical effects of their more massive parent. To make progress, it is necessary to augment and solidify the census of dwarf satellites of galaxies outside the Local Group. M\"uller et al. (2015) presented 16 dwarf galaxy candidates near M83, but lacking reliable distances, it is unclear which candidates are M83 satellites. Using red giant branch stars from the HST/GHOSTS survey in conjunction with ground-based images from VLT/VIMOS, we confirm that one of the candidates, dw1335-29 --- with a projected distance of 26\,kpc from M83 and a distance modulus of $(m-M)_0=28.5^{+0.3}_{-0.1}$ --- is a satellite of M83. We estimate an absolute magnitude $M_V=-10.1\pm{0.4}$, an ellipticity of $0.40^{+0.14}_{-0.22}$, a half light radius of $656^{+121}_{-170}$ pc, and [Fe/H] = $-1.3^{+0.3}_{-0.4}$. Owing to dw1335-29's somewhat irregular shape and possible young stars, we classify this galaxy as a dwarf irregular or transition dwarf. This is curious, as with a projected distance of 26\,kpc from M83, dw1335-29 is expected to lack recent star formation. Further study of M83's dwarf population will reveal if star formation in its satellites is commonplace (suggesting a lack of a hot gas envelope for M83 that would quench star formation) or rare (suggesting that dw1335-29 has a larger M83-centric distance, and is fortuitously projected to small radii).
\end{abstract}



\section{Introduction}


In the $\Lambda$CDM paradigm, the galaxies that we see are condensations of cold gas and stars in the centres of more extended and more massive dark matter haloes \citep{white78}. Dark matter haloes assemble in great part from the merging of smaller haloes at earlier time; accordingly, dark matter halos should have a large number of smaller `sub-haloes' - concentrations of dark matter embedded in the larger halo. Around a galaxy like the Milky Way, many of these sub-halos might have been expected to host visible dwarf galaxies \citep{moore99}. In the Local Group, there are many fewer dwarf galaxies at a given velocity scale than dark matter sub-haloes \citep{moore99,klypin99}, a problem termed the `missing satellites problem'. While the discovery of many more Local Group satellites has helped enrich and re-frame this problem (e.g., \citealp{martin13,laevens15,belokurov07,boylankolchin11,kimjerjen15}), and recent simulations appear to successfully reproduce the numbers and properties of satellites by including reionization and stellar feedback effects e.g. \citep{sawala16,wetzel16}, the distribution, number, and properties of low-mass galaxies offer crucial insight into the formation and evolution of low-mass dark matter haloes and the galaxies that live in them. 

The ongoing search for Local Group dwarfs has also brought into sharper focus the apparently planar distribution of Milky Way \citep{pawlowski12} and M31 \citep{koch06} satellites. Like the missing satellites problem, this planar distribution appears at first sight challenging to interpret in a $\Lambda$CDM (\citealp{pawlowski14}, \citealp{pawlowski15}; although see, e.g., \citealp{cautun15} for arguments that such planes should be uncommon but should exist), and could give important insight into our understanding of galaxy formation in a cosmological context (e.g., \citealp{libeskind15}). Extending the satellite search and characterization beyond the Local Group \citep{muller16} is therefore necessary to have a better sense of how widespread such alignments are in a more statistical sample. 

One key limitation of current datasets is that a detailed census of satellite galaxies exists only around the Milky Way and M31. Faint satellite galaxies are diffuse and have low surface brightness (SB), so much so that discovering them in the Local Group is still an ongoing effort  (e.g., \citealp{koposov15,bechtol15}). It is unknown how important the limitation of our satellite census to the Local Group is. Do the Milky Way and M31 have somewhat lower mass than expected? If so, this dramatically lessens tensions between predictions of large numbers of massive dwarfs and the modest number of observed dwarf galaxies at such velocity scales (the evidence for an uncertain Milky Way mass and its consequences are discussed by e.g., \citealp{wang12,vera-ciro13}). Do dwarf galaxies near their primary tend to lack on-going star formation in general (e.g., \citealp{grebel03,slater14,grcevichputman09})? If so, this may suggest that hot gas haloes around roughly Milky Way mass primaries are ubiquitous (e.g., \citealp{mayer06,mayer01,gatto13,slater13}). If instead many proximate dwarf satellites host significant star formation (as may be seen around M106 by \citealp{spencer14}),  
this may suggest considerable halo-to-halo scatter in hot gas content or dwarf accretion time.

Owing to their low SB and the necessity of covering large areas of sky to significant depth, the census of dwarf galaxies around nearby massive hosts is still relatively incomplete. Search methods are varied, from doing blind H{\sc i} surveys (e.g. HIPASS, \citealp{barnes01}), to targeting low velocity H{\sc i} detections in the field (e.g., \citealp{bernsteincooper13}), to searching for faint low SB candidates \citep{karachentsev07,chiboucas13,spencer14,muller15,merritt14}, 
to discovering dwarf satellites by virtue of their being resolved into individual stars, enabling a distance measurement \citep{monachesi14,crnojevic16}. The discovery of low SB candidates is a powerful and relatively inexpensive method of finding candidates (e.g., \citealp{chiboucas09, merritt14}); yet, distances from resolved stellar populations \citep{chiboucas13} or SB fluctuations \citep{jerjen01,rekola05}, or redshifts from follow up spectroscopy \citep{spencer14} are required to confirm that the candidates are in fact nearby dwarf galaxies instead of unresolved background low SB galaxies. 


We have used the Hubble Space Telescope (HST) and the Very Large Telescope (VLT) to characterize the newly discovered M83/NGC 5236 dwarf satellite candidate dw1335-29 \citep{muller15}. This is a step towards characterizing faint galaxy populations around nearby galaxies outside the Local Group. Furthermore, studying this dwarf galaxy in detail aided by the resolved stellar population, and in the context of previously discovered CenA/M83 dwarfs \citep{cote97,banks99,jerjen00,crnojevic14,crnojevic16}, gives a better picture of M83's environment. In \S 2 we introduce the data, their reduction, and photometry; \S 3 focuses on the dwarf's properties and how they were measured, \S 4 discusses the results and possible implications of the properties of dw1335-29, and we conclude in \S 5.

\section{The Data, Data Reduction and Photometry}

\citet{muller15} detected 16 extended low SB candidate M83 companions using data from the Dark Energy Camera (DECam) at the CTIO 4 m Blanco Telescope. Yet, without resolving those candidates into stars from which the tip of the red giant branch (TRGB) could be determined, there is no distance information available. It is unclear if these low SB objects are indeed dwarf companions of M83 or are more distant background galaxies close to M83 in projection; consequently, their properties are highly uncertain. 

We had independently discovered one of the dwarf candidates, dw1335-29, as an overdensity of resolved red giant and young stars in one of the fields (M83 Field 10) of the GHOSTS (Galaxy Halos, Outer disks, Substructure, Thick disks and Star clusters) survey (\citealp{radburnsmith11}, \citealp{monachesi16}; PI: R.\ de Jong, HST programs = 10523, 10889, 11613, and 12213). This particular pointing along with three others (PI: D.\ A.\ Thilker, HST program = 10608) were taken to follow up on the extended UV disk detected by GALEX \citep{thilker05}. Field 10's F606W and F814W total exposure times are 1190s and 890s respectively, split between two exposures per filter. The images were combined, extended sources were determined and removed, and stars were identified following \citet{radburnsmith11}. In addition, artificial star tests (ASTs) were performed to assess the completeness of our data as well as the photometric uncertainties \citep{radburnsmith11}. Fake stars with realistic magnitudes and colors are added onto the field and photometry is performed again. The ratio between recovered and injected fake stars for a given colour-magnitude bin determines the completeness of our data. Differences between the magnitudes and colours of recovered and injected fake stars are used to estimate the photometric uncertainty. An overdensity of primarily red giant branch-coloured stars with a modest overdensity of blue stars was seen at the position of dw1335-29, confirming it to be a nearby dwarf galaxy and permitting a distance estimate and surface brightness profile measurement using its red giant branch (RGB) stars.


However, the dwarf was only partially covered by GHOSTS and so the data were insufficient to characterize its total properties --- total brightness, position, ellipticity and half-light radius. Accordingly, we use imaging data from Very Large Telescope's (VLT) Visible MultiObject Spectrograph (VIMOS; Proposal ID: 386.B-0788(B), PI: R.\ de Jong) to globally characterize dw1335-29. Sixteen images with the V band filter totaling an exposure time of 4800 s and 7 images with the I band filter totaling 2100 s were taken on 5 March 2011. These VIMOS data consist of 4 chips with 7'$\times$8' field of view each. For our current purposes, we consider only the 6.67' x 3.76' region around  dw1335-29 (9.7 kpc $\times$ 5.5 kpc at the adopted distance of dw1335-29), which is large enough to include a significant region around the dwarf but small enough that point spread function (PSF) variations across the frame were negligible.

\begin{figure}
	\includegraphics[width=\columnwidth]{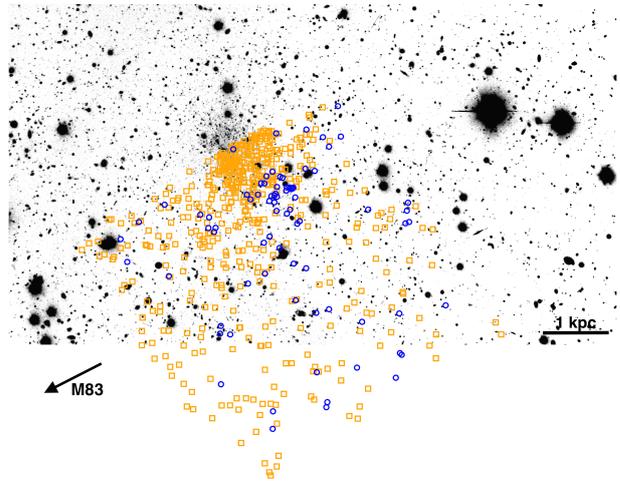}
    \caption{A VIMOS $V$-band image of dw1335-29. The image has a field size of 6.67'$\times$3.76' (9.7 kpc $\times$ 5.5 kpc at the adopted distance of dw1335-29). Point source detections from the GHOSTS survey are overlaid. Orange squares denote point sources with the magnitudes and colour of RGB stars at the distance of M83, whereas blue open circles correspond to point sources with the magnitudes and colours consistent with upper main sequence stars (following Fig.\protect\ref{fig:cmd}). East is to the left and north is up. The dwarf lies 26 projected kpc from M83, and with a distance modulus of $m - M = 28.5^{+0.3}_{-0.1}$, is likely a member of the M83 group.}
    \label{fig:field}
\end{figure}

\begin{figure}
	\includegraphics[width=\columnwidth]{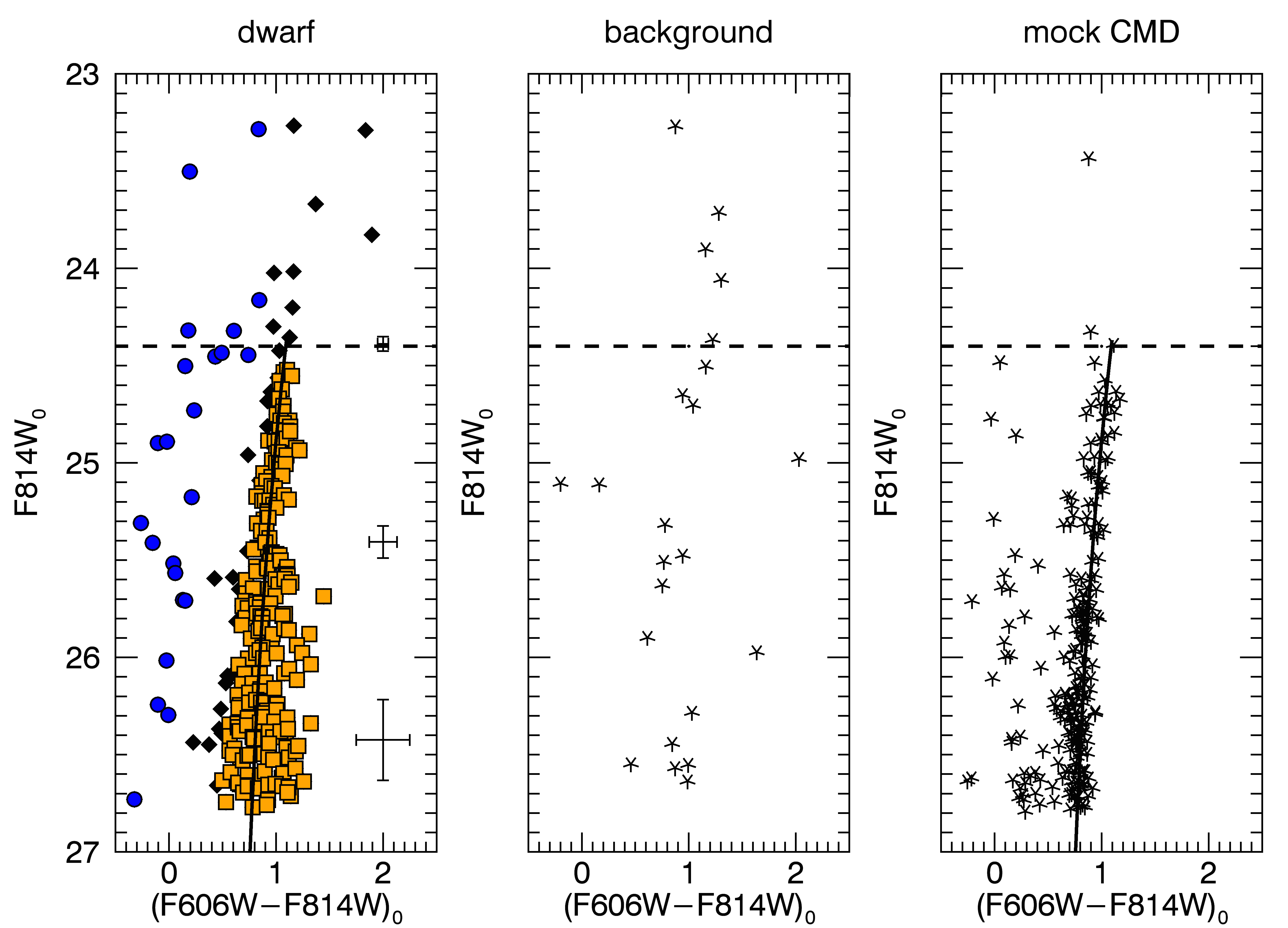}
	\includegraphics[width=\columnwidth]{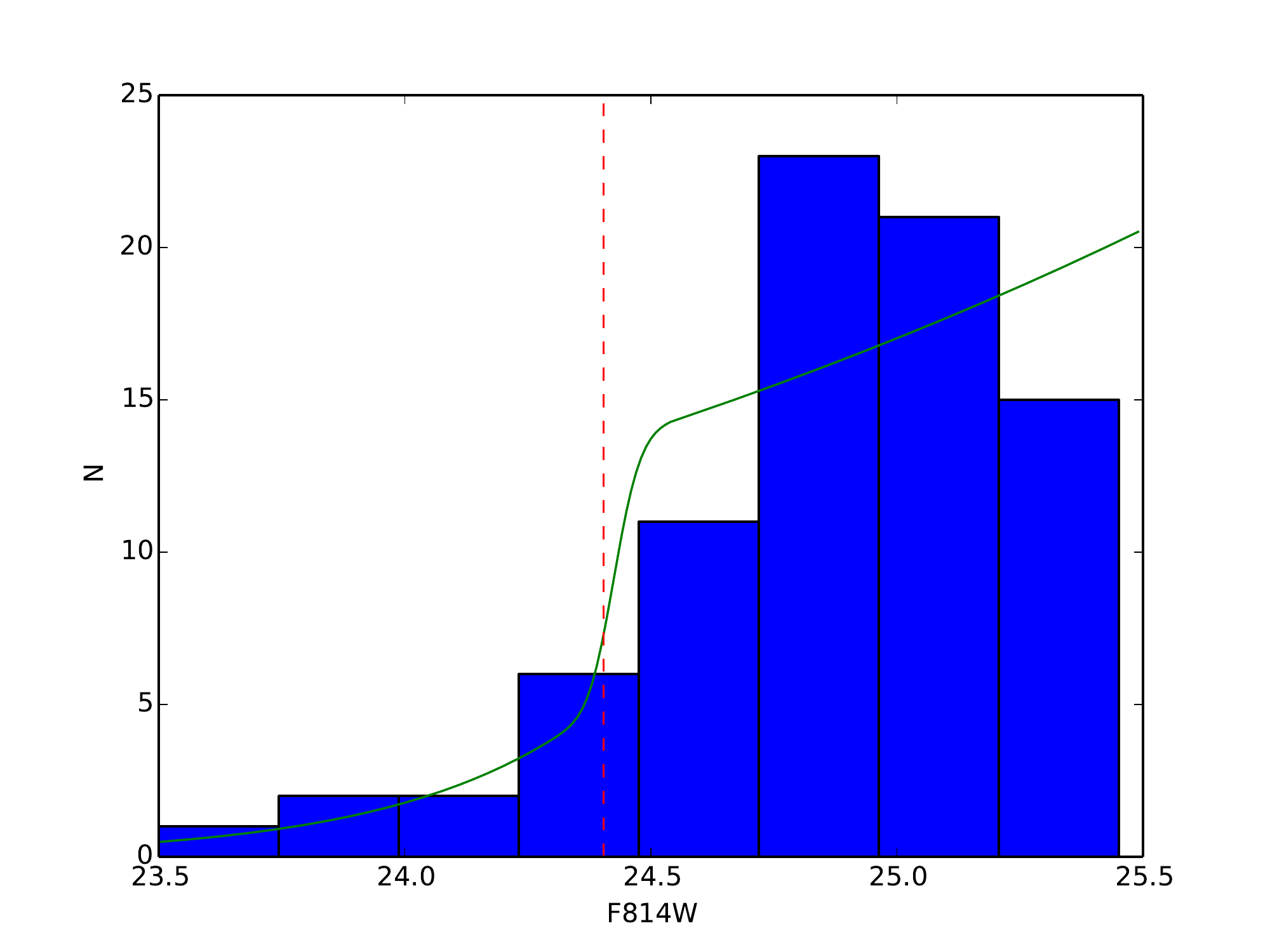}
    \caption{Foreground extinction-corrected CMD and LF of dw1335-29. Top left: Stars within twice the half-light radius of dw1335-29 and with magnitudes and colours consistent with being RGB stars at the distance of M83 are shown as orange squares, upper main sequence stars as blue circles, and black diamonds for all other stars. The best fit isochrone with [Fe/H] = $-1.3^{+0.3}_{-0.4}$ at 10 Gyr is overlaid as the solid line (in first and third panels) and the TRGB overlaid as the dashed black line (same for all panels). The photometric uncertainties as a function of magnitude at colour=1, as derived from ASTs, are also shown. Top center: A CMD of a different part of the field but with the same area as the dwarf, showing that it is less dense and there are many fewer blue stars. Top right: A mock CMD with constant star formation history, [Fe/H] = $-1.3$ and $M_V \sim -10$, convolved with expected photometric error. A constant star formation history gives a very similar number of RGB and bluer stars to the observations (left). Bottom panel: LF of RGB stars in dw1335-29, where the TRGB is estimated to have $m_{\rm F814W} =24.40^{+0.3}_{-0.1}$, marked as the red dashed line. The green solid line shows the best-fit analytic LF smoothed by the magnitude error.} 
    \label{fig:cmd}
\end{figure}

The images were bias subtracted, flat fielded, aligned and stacked into combined $V$-band and $I$-band images using standard IRAF routines. The resulting deeper VIMOS $V$-band image is shown in Fig. \ref{fig:field} overlaid with stellar detections from GHOSTS (open circles and squares). Detections are colour coded by their position on the colour--magnitude diagram (CMD) --- orange squares are consistent with RGB stars at the distance of dw1335-29, whereas blue open circles are consistent with younger blue stars at that distance. For comparison, we plot the CMD of the dwarf side by side with a CMD of a field of the same size (see top left and top center panels of Fig. \ref{fig:cmd}) to show that there is a sizable difference in the concentration of stars. Interestingly, the blue stars in Fig. \ref{fig:field} are concentrated off-centre towards the southwest direction, indicating a region of star formation.

dw1335-29 is relatively compact, and its central regions are crowded in the VIMOS imaging data. Furthermore, VIMOS's PSF is large enough to lead to contamination of `point source' catalogs with background galaxies. Accordingly, we performed both point source photometry for the VIMOS data, to calibrate the VIMOS data using the GHOSTS star catalog, and surface photometry, to derive the global properties of dw1335-29. Point sources were detected in the deeper VIMOS $V$-band images and forced photometry was performed on the $I$-band data. Stars in common with GHOSTS were identified by cross-matching their RA and Dec up to 0.5" accuracy, and theoretical isochrones from \citet{marigo08} and \citet{girardi10} were used to predict the $V$ and $I$-band magnitudes of HST-detected stars given their F606W and F814W magnitudes. We estimate that the final calibration is accurate to $\sim$10\% in each band, with dominant sources of error being our uncertainty in HST calibration (\citealp{radburnsmith11} find that stars in overlapping GHOSTS fields have photometry that repeats to $\sim 5$\% accuracy). The $V$ and $I$ values derived from the isochrone agree with the $V$ and $I$ transformations \citep{sirianni05} for F606W and F814W, respectively. The data were then corrected for foreground Galactic extinction following  \citet{schlafly11}.

\begin{figure}
	\includegraphics[width=0.49\columnwidth]{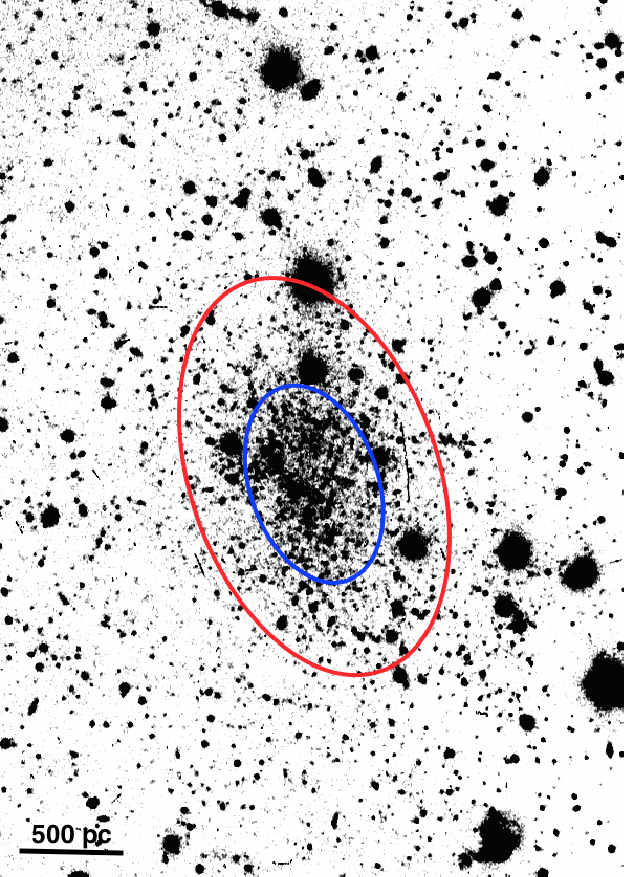}
	\includegraphics[width=0.49\columnwidth]{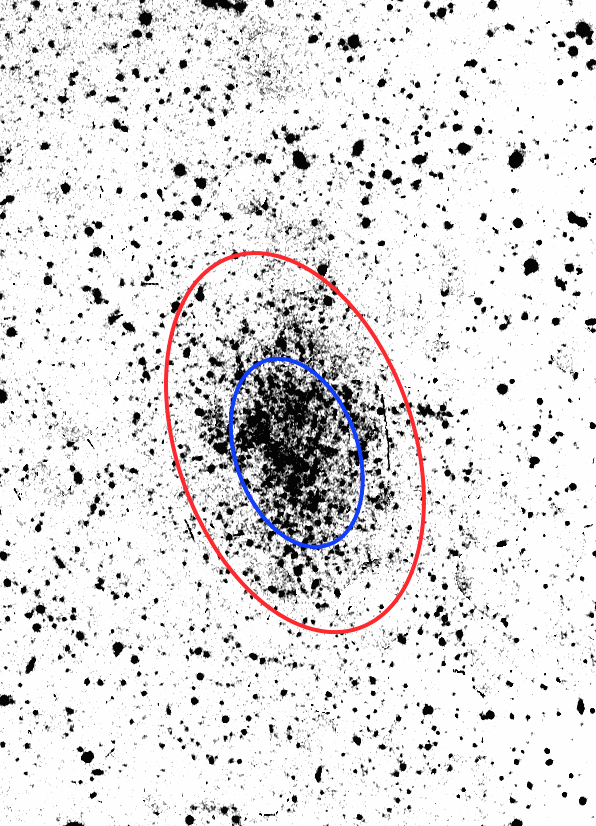}
	\includegraphics[width=\columnwidth]{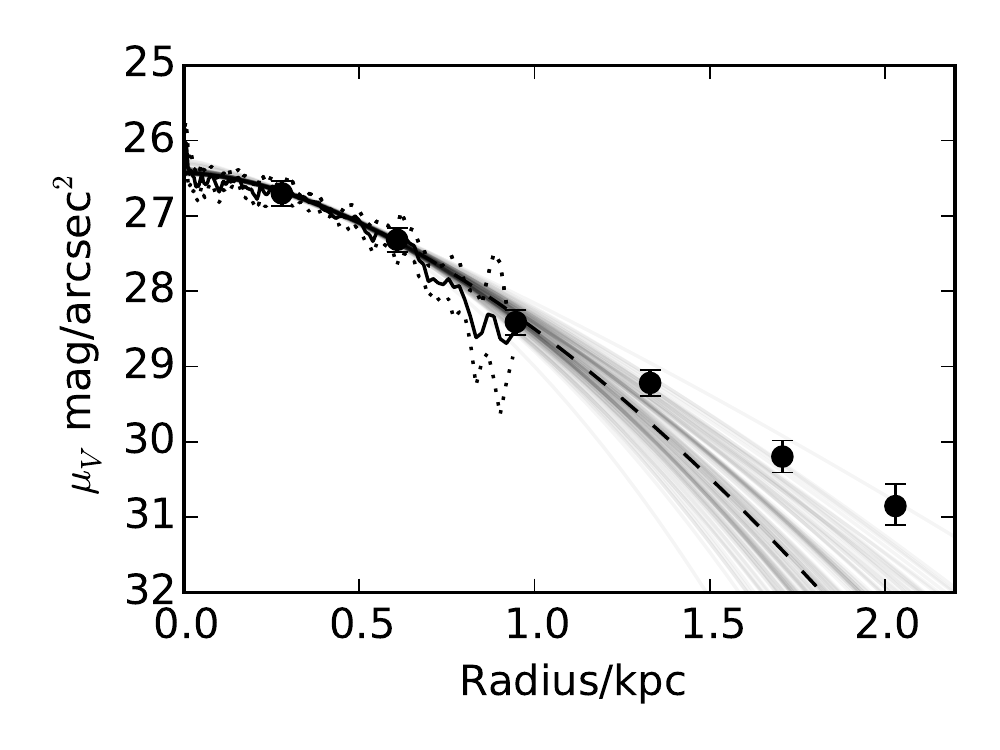}
    	\caption{Top: An enlarged view of the VIMOS $V$-band image of dw1335-29 with ellipses overlaid corresponding to one (inner blue) and two half-light radii (outer red). The left panel shows the data before masking the bright objects while the right panel shows the image with masked out bright foreground stars and background galaxies. Bottom: $V$-band SB profile of the dwarf using VIMOS $V$-band surface photometry (solid line for the median value and dotted black lines for $\pm$ sigma uncertainties), estimated $V$-band flux from scaling background contamination-subtracted RGB stars counts using the HST/GHOSTS data (circles), and the best S\'ersic profile fit to the combined SB profile (black dashed line, with 100 bootstrap resampled fits shown in light gray to give an indication of the uncertainty in the fits). The best-fit parameters are $M_V=-10.1\pm{0.4}$, $r_h = 656^{+121}_{-170}$ pc along the semi-major axis, and $n = 0.61^{+0.13}_{-0.26}$.}
    	\label{fig:ellipse}
\end{figure}

In order to estimate the global properties of dw1335-29, we use IRAF's STSDAS/ELLIPSE package to perform surface photometry on the deeper combined VIMOS $V$ band image. Bright foreground stars and background galaxies were first replaced with the average sky value and noise in their local region. Both the original image and image with bright stars and background galaxies masked out are shown in Fig. \ref{fig:ellipse} on the top left and right panels, respectively. dw1335-29 has a low central SB $\mu_V \sim 26.5$ mag/arcsec$^2$ (as is clear in the lower panel in Fig. \ref{fig:ellipse}), leading to considerable uncertainty in the galaxy's centre, position angle and ellipticity. Furthermore, in such cases, ELLIPSE often fails to converge on a reasonable answer. Accordingly, we run ELLIPSE a number of times with different trial values of central position, ellipticity and position angle (keeping some values fixed while leaving others to float) to give a sense of the reasonable range for each value and give a range of values for derived half-light radius and total brightness.  

We extend the surface brightness profile of dw1335-29 to substantially larger radii and lower surface brightness (equivalent to $\mu_V \sim 31$ mag/arcsec$^2$) by using the number of RGB stars per unit area to estimate its $V$-band surface brightness (see also \citealp{Harmsen16}). We count the number of RGB stars with $24.4<{\rm F814W}<26.4$ per unit area along elliptical annuli defined from the fits to the VIMOS data (as the HST data alone cannot determine the centre, ellipticity and position angle). We correct for completeness using the results from the ASTs. We then subtract off a constant background value (a combination of diffuse M83 halo stars, Milky Way foreground stars and unresolved galaxies), determined using the most distant annuli. In order to scale the RGB star counts to $V$-band surface brightness, we calculate the $V$-band luminosity per RGB star assuming a constant SFH  (see \S \ref{sec:disc} for discussion of this assumption) and adopting the best fit metallicity with [Fe/H] $= -$1.3. Where the VIMOS surface photometry and star count-derived profile overlap, they agree to within their uncertainties.




\begin{table}
	\caption{Properties of dw1335-29}
	\begin{threeparttable}
	\centering
	\label{tab:example_table}
	\begin{tabular}{|c|c|}
		\hline
		Parameter & Value\\
		\hline
		$\alpha$ (FK5) & 13\textsuperscript{h}35\textsuperscript{m}46\textsuperscript{s}$.9^{+0.1}_{-0.3}$\\
		$\delta$ (FK5) & $-29^{\circ}42'22''.4^{+1.8}_{-6.7}$\\
		E(B - V)\tnote{a} & 0.04\\
		projected D[kpc] from M83 & 26 \\
		$M_V$ & {\bf $-10.1\pm0.4$}\\
		\textit{(m - M)}\textsubscript{0} & $28.5^{+0.3}_{-0.1}$\\
		Ellipticity & $0.40^{+0.14}_{-0.22}$\\
		Position angle (N to E) & $19^{\circ +8}_{-17}$\\
		\textit{r\textsubscript{h}}[arcsec] & {\bf $27^{+5}_{-7}$}\\
		\textit{r\textsubscript{h}}[pc] & {\bf $656^{+121}_{-170}$}\\
        S\'ersic index $n$ & {\bf $0.61^{+0.13}_{-0.26}$}\\
      	         \text{[Fe/H]} & $-1.3^{+0.3}_{-0.4}$\\
		\hline
		\hline
		\end{tabular}
	\begin{tablenotes}
      		\item[a] \citet{schlegel98}
   	 \end{tablenotes}
	 \end{threeparttable}
\end{table}

\section{Properties}

Given the range of reasonable ELLIPSE fits, we estimate that dw1335-29's centre is located at FK5 ($\alpha$,$\delta$) (13\textsuperscript{h}35\textsuperscript{m}46\textsuperscript{s}$.9^{+0.1}_{-0.3}$,$-29^{\circ}42'22''.4^{+1.8}_{-6.7}$), with ellipticity and position angle of $0.40^{+0.14}_{-0.22}$ and 19$^{\circ +8}_{-17}$ respectively. In comparison, \citet{muller15} measured the centre coordinates to be J2000($\alpha$,$\delta$) (13\textsuperscript{h}35\textsuperscript{m}46\textsuperscript{s},$-29^{\circ}42'24''$). Our value for ellipticity, while being quite uncertain, is reasonable for dwarf galaxies of this luminosity \citep{martinetal08,mcconnachie12}.

In order to estimate the metallicity and distance of dw1335-29, we use the RGB stars from GHOSTS (top-left panel of Fig. \ref{fig:cmd}). We compare the CMD of stars within 2 half light radii of dw1335-29, limited to the upper 1.5 mag, to isochrones \citep{marigo08,girardi10} with a range of metallicities from Z=0.0001 ([Fe/H] = $-2.28$) to Z=0.0021 ([Fe/H] $= -0.96$). Minimizing chi-square gives an acceptable match for isochrones with [Fe/H] $\sim -1.3^{+0.3}_{-0.4}$, where uncertainties are derived from bootstrapping the RGB stars, accounting for systematic errors from photometric calibration and isochrone uncertainties \citep{streich14}. Selecting stars in a broad range of colour centred on this isochrone, we construct the luminosity function (LF) of the RGB stars (lower panel of Fig. \ref{fig:cmd}). In order to locate the TRGB, we follow \citet[their Appendix C]{monachesi16} by maximum-likelihood modeling the LF as two power laws separated by a discontinuous jump at the TRGB, and then finding the value of the TRGB magnitude and power law slopes that maximize the likelihood that the observed RGB stars are drawn from the model LF convolved with the photometric errors (the green line in bottom panel of Fig. \ref{fig:cmd}). We find the TRGB to be located at $m_{F814W} = 24.4^{+0.3}_{-0.1}$. In order to calculate the expected absolute magnitude, we adopt Equation C1 from \citet{monachesi16}: 
\begin{equation}
M_{F814W}=-4.06+0.20[(F606W-F814W)-1.23],
\end{equation}
 where $F606W-F814W$ denotes the mean colour of stars within 0.2 mag of the TRGB (determined to be $\sim1$ by fitting a Gaussian), giving us a distance modulus of $(m-M)_0 = 28.5^{+0.3}_{-0.1}$, where the errors are determined by bootstrap resampling. Note that the maximum likelihood method operates on the underlying photometric data rather than the histogram, and so is not biased by chance alignments with bin edges (such as the one that appears as a jump at F814W=24.7). This places dw1335-29 at the same distance as M83 to within its uncertainties i.e. $(m-M)_0 = 28.41\pm0.1$ from TRGB \citep{monachesi16}, $(m-M)_0 = 28.25\pm0.15$ from Cepheids \citep{thim03}, and at a similar distance to other M83 companions (at larger projected radii; \citealp{karachentsev05}).

To determine the absolute magnitude and half-light radius of dw1335-29, we combine the VIMOS $V$-band surface brightness profile from the ELLIPSE fits to the unresolved light of dw1335-29 (shown as solid and dotted black lines in Fig. \ref{fig:ellipse}) with the $V$-band surface brightness profile inferred from the GHOSTS RGB star counts (shown as filled circles) that were background contamination-subtracted. The $V$-band surface brightness profile and inferred parameter uncertainties include an estimate of systematic uncertainties from foreground star subtraction, ranging from relatively minimal subtraction to very aggressive subtraction. We fit the combination of the integrated light and star-count derived surface brightness profile with a S\'ersic profile of the form $SB \propto e^{-r^{1/n}}$ \citep{sersic63, graham05}. The SB profile and best fit S\'ersic profile with bootstrap uncertainties are shown in Fig. \ref{fig:ellipse}, where the dashed black line shows the best fit and and the gray lines show the 100 bootstraps. The parameters and formal uncertainties for the single S\'ersic profile fit, accounting for ellipticity and distance uncertainties, are : $m_V = 18.47\pm0.15$ and $M_V = -10.1\pm{0.4}$ for total apparent and absolute magnitude, half light radius \textit{r\textsubscript{h}}=$27^{+5}_{-7}$ arcsec or $r_h = 656^{+121}_{-170}$\,pc along the semi-major axis and S\'ersic index $n = 0.61^{+0.13}_{-0.26}$. The luminosity is obtained by integrating the combined SB profile, extrapolating using the S\'ersic fits. We also note that a single S\'ersic profile is not a particularly accurate description of the observed surface brightness profile, especially at $r>1.5$\,kpc. $M_V$ and $r_h$ measured using this method are within the uncertainties of estimates calculated from the VIMOS data only, but have improved uncertainties owing to better knowledge of the surface brightness profile at large radii and faint limits. All the dwarf's derived properties are listed in Table 1.

\begin{figure}
	\includegraphics[width=\columnwidth]{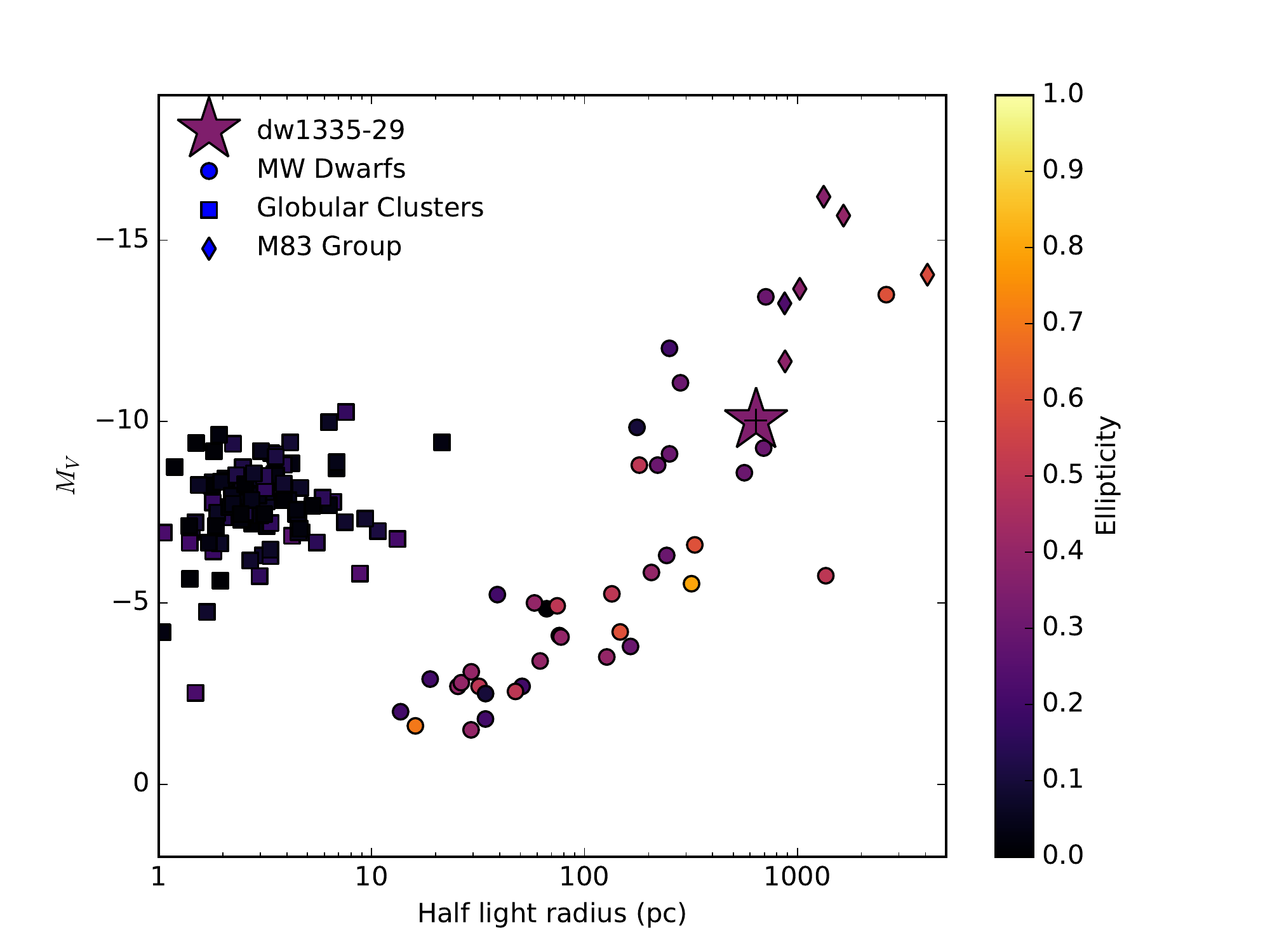}

    	\caption{$M_V$ vs $r_h$ of MW satellites (circles; \citealp{mcconnachie12}), MW globular clusters (squares; \citealp{harris10}), known M83 group satellites (diamonds; \citealp{karachentsev05}), and dw1335-29 (star). The objects are colour-coded by their ellipticities.}
    	\label{fig:Mv_rh}
\end{figure}

\begin{figure}
	\includegraphics[width=\columnwidth]{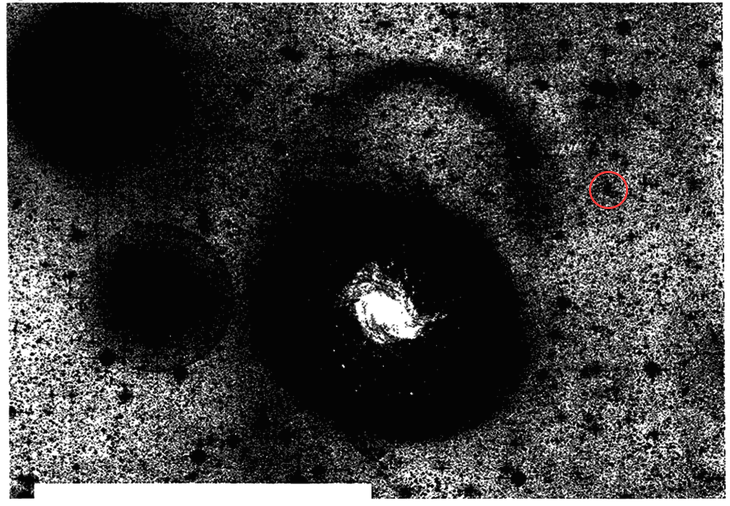}
    	\caption{A deep photographic image of M83 taken using the UK Schmidt Telescope by \citet[figure reproduced with permission]{malin97}, as shown in their Fig.\ 5 where the white scale bar corresponds to 30 arcmin. The northern stream appears not to be associated with dw1335-29, circled in red.}
    	\label{fig:Malin}
\end{figure}

\section{Discussion} \label{sec:disc}


Do dw1335-29's properties fit in with known dwarf galaxies and satellites of M83? In Figure \ref{fig:Mv_rh}, we show the distribution of half light radii, absolute magnitudes and ellipticities of Milky Way satellite galaxies (circles), globular clusters (squares), M83 satellites (diamonds), with the properties of dw1335-29 shown as a star. dw1335-29 falls squarely in the range of properties of dwarf satellites of the Milky Way, and while it is the faintest confirmed member of the M83 group, it falls along an extrapolation of the relationship between their sizes and luminosities. 

Given that M83 has both a prominent stellar tidal stream dominated by old stellar populations \citep{dejong07,barnes14} and an extended HI disk \citep{huchtmeier81,miller09}, it is worth briefly exploring whether dw1335-29 is related to either structure. dw1335-29 is several arcminutes (meaning several projected kpc) from the stellar tidal stream north of M83, shown in Fig.\ \ref{fig:Malin} from \citet{malin97}. The dwarf has no indication of a connection to the stream in the VIMOS or HST imaging, or in more recent unpublished Subaru imaging covering the whole area. We conclude that dw1335-29 is extremely unlikely to be related to M83's prominent stellar tidal stream. 

dw1335-29's position is projected towards the outskirts of M83's extended H{\sc i} and UV disk (\citealp{bigiel10,dong08,thilker05}). Given that dw1335-29 is dominated by an old metal-poor stellar population and has a luminosity, half-light radius and ellipticity that place it squarely on the scaling relations obeyed by Local Group galaxies, we consider it extremely unlikely that the main stellar body of dw1335-29 is associated with M83's H{\sc i}/UV disk; we consider this to be a coincidental projection.


While there are blue stars across the full extent of the GHOSTS field (and many other GHOSTS fields within 25kpc of M83; \url{http://vo.aip.de/ghosts/}), and it is possible that the concentration of blue stars within 500 projected pc of the dwarf in Fig.\ \ref{fig:field} are projections, we argue against that possibility for two reasons. First, while there are many concentrations of star formation on the southern side (and less so on the northern side) of M83 evident in deep GALEX imaging \citep{thilker05}, there are {\it none} on the north-western side at distances $>$15\,kpc from M83 apart from dw1335-29. Given that there is only one visible knot of star formation on the north-western side (that associated with dw1335-29) and that it is within 500pc of the centre of the RGB star concentration of dw1335-29, a very crude estimate of the probability of a chance projection in that $\sim 20$\,kpc square region is {\bf $\sim \pi(\,0.5\,{\rm kpc})^2 / (20\,{\rm kpc})^2 \sim 0.2$\%.} Second, dw1335-29's old stellar populations appear rather irregular both in Figs.\ \ref{fig:field} and \ref{fig:ellipse}, lending support to the notion that the young stars projected near dw1335-29 belong to it. Given these apparent coincidence between dw1335-29 and the young stars on one hand, and the slightly irregular appearance of dw1335-29 on the other, we tentatively classify it as a dwarf irregular (dIrr; given its ongoing star formation) or transition dwarf (dTr; given that its appearance is dominated by old stellar populations; \citealp{karachentsev14}). In order to explore this idea more quantitatively, we simulated dw1335-29's CMD by assuming a constant star formation history, $M_V \sim -10$ and [Fe/H]$=-1.3$, placing the simulated stellar population at the distance of dw1335-29. We show a mock CMD in the top-right panel of Fig. \ref{fig:cmd}, finding a similar ratio of blue to RGB stars in the simulated CMDs as in the observations. Many dwarf galaxies have approximately constant SFHs (e.g., \citealp{weisz14}), lending weight to the idea that dw1335-29 should be classified as a dwarf irregular or transitional dwarf.  Given its absolute magnitude and half-light radius, dw1335-29 appears to have similar properties to Local Group transition dwarfs LGS \citep{miller01} and Phoenix dwarf galaxy \citep{hidalgo09}. Unlike the off-centre star formation in dw1335-29, both Local Group analogs have centrally concentrated star-formation regions. The explanation for this difference (if one is required) is unclear; this may be a sign that the star-forming gas is in the process of being stripped from dw1335-29. 

Assuming that dw1335-29 has ongoing star formation, it is an interesting question to consider how a dwarf with a projected separation of only 26 kpc from M83 can host star formation. In the Local Group, galaxies comparable to dw1335-29 in mass are devoid of star formation if they are within 100 kpc of the primary galaxy; indeed, some galaxies lack star formation up to 700 kpc or so from their nearest large neighbor (e.g., \citealp{grebel03,mcconnachie12}). This is most naturally interpreted as indicating that dwarf satellites are highly susceptible to loss of their cold gas content through tidal or gas dynamical interactions with their primary galaxy, with even one pass within 100 kpc of a larger primary being sufficient to lead to the loss of its cold gas (e.g., \citealt{mayer01,mayer06}; \citealp{grebel03}; \citealp{slater13}). If indeed (as we argue) dw1335-29 forms stars, it suggests one of two possibilities: either dw1335-29 has a larger M83-centric distance than its projected distance, or M83 lacks the tidal/gas dynamical environment necessary to quench star formation. In support of a larger distance for dw1335-29, we note that dw1335-29 appears to lack prominent tidal features (although the off-centre star-forming region may suggest mild gas stripping) and that many dwarf galaxies within 30 projected kpc of their primary are being tidally disrupted by the primary (e.g., Sagittarius; \citealp{johnston95}, and NGC 253's disrupting companion with a projected separation of 50kpc; \citealp{toloba16}). On the other hand, \citealp{karachentsev02}'s survey of M83 group galaxies has 7/10 of them forming stars, consistent with the possibility of a reduced efficiency for tidal and gas dynamical stripping from M83. A comprehensive census of M83 satellites and their star formation properties would help to resolve this issue.

\section{Conclusions}

Using a combination of VIMOS ground-based data and HST resolved stellar imaging, we have confirmed that the dwarf candidate dw1335-29 is indeed a satellite of M83 with a TRGB-derived distance modulus of $(m-M)_0$ = $28.5^{+0.3}_{-0.1}$. By combining unresolved VIMOS light with HST star counts, we construct a surface brightness profile that reaches an equivalent $V$-band surface brightness of $\sim 31$ mag/arcsec$^2$. With an absolute magnitude $M_V=-10.1\pm{0.4}$, major axis half-light radius $r_h = 656^{+121}_{-170}$ pc, S\'ersic index $n = 0.61^{+0.13}_{-0.26}$ and ellipticity of $0.40^{+0.14}_{-0.22}$, dw1335-29 falls on the relationships between absolute magnitude, size and ellipticity typical of Local Group and known M83 dwarf satellites. Due to the likely presence of a star-forming region in the galaxy, we classify it either as an irregular dwarf or a transition dwarf. With a projected distance of 26 kpc from M83, it is mildly surprising that dw1335-29 has ongoing star formation; perhaps dw1335-29 has a larger M83-centric distance and is coincidentally projected close to its host, or perhaps M83 lacks a sufficiently dense hot gas envelope to quench star formation in its satellites (this would appear plausible if future studies show that most of M83's satellites have ongoing star formation). 

\section*{Acknowledgments}

This work was partially supported by
NSF grant AST 1008342 and HST grants GO-11613 and GO-12213 provided by NASA through a grant from the Space Telescope Science Institute, which is operated by the Association
of Universities for Research in Astronomy, Inc., under NASA contract NAS5-26555.
This research has made use of NASA's Astrophysics Data System Bibliographic Services.






\bibliographystyle{mnras}
\bibliography{dwarf}






\label{lastpage}
\end{document}